\documentclass[preprint,aps,superscriptaddress,nofootinbib]{revtex4-1}
\usepackage{graphicx}
\usepackage[toc,page]{appendix}
\usepackage{braket}
\newcommand{\del}{\partial}

\begin{document}
\title{A degenerate extension of the Schwarzschild exterior}

\author{Romesh Kaul}
\email{kaul@imsc.res.in}
\affiliation{The Institute of Mathematical Sciences, Chennai-600113, INDIA}

\author{Sandipan Sengupta}
\email{sandipan@phy.iitkgp.ernet.in}
\affiliation{Department of Physics and Centre for Theoretical Studies, Indian Institute of Technology Kharagpur, Kharagpur-721302, INDIA}

\begin{abstract}

We present vacuum spacetime solutions of first order gravity, which are described by the exterior Schwarzschild geometry in one region and by degenerate tetrads in the other. The invertible and noninvertible phases of the tetrad meet at an intermediate boundary across which the components of the metric, affine connection and field-strength are all continuous. Within the degenerate spacetime region, the noninvertibility of the tetrad leads to nonvanishing torsion. In contrast to the Schwarzschild
spacetime which is the unique spherically symmetric solution of Einsteinian gravity, all the field-strength components associated with these vacuum geometries remain finite everywhere.

%

\end{abstract}
  
\maketitle
\section{Introduction}
In Einstein's theory of gravity in vacuum, the Schwarzschild metric turns 
out to be the unique spherically symmetric solution. This geometry 
exhibits a curvature singularity at the origin.
As long as the metric is demanded to be invertible and spherically 
symmetric, there seems to be no escape from such singular solutions 
in the classical theory. 
 However, the first order gravity theory in vacuum  admits, besides 
a phase with invertible tetrads (metric), another (non-Einsteinian) 
phase based on  tetrads which have vanishing determinants and hence 
 are not invertible. The classical theories for these two phases are not 
equivalent \cite{tseytlin}. In fact, the solution space with noninvertible 
tetrads possesses a rich structure, as was elucidated in some recent 
studies \cite{kaul,kaul1}. In view of this, it is worthwhile to explore 
whether there could be any extension of the Schwarzschild exterior 
geometry such that the full spacetime is regular everywhere, within 
a formulation of gravity theory that admits both the phases. 

To deal with degenerate spacetime solutions in gravity theory, the 
appropriate starting point is  the first order formulation based on
Hilbert-Palatini action, which, unlike the second order formulation, 
does not require the explicit use of the inverse metric. This  action 
functional is given in terms of two independent fields, the tetrad 
$e_\mu^I$ and the connection $\omega_\mu^{IJ}$, as:
\begin{eqnarray}\label{HP}
S&=&\frac{1}{8\kappa^2}\int d^4 
x~\epsilon^{\mu\nu\alpha\beta}\epsilon^{}_{IJKL}
e_{\mu}^I e_\nu^J R_{\alpha \beta}^{~~KL}(\omega)
\end{eqnarray}
Here $R_{\mu \nu}^{~~IJ}(\omega)=\del_{[\mu}\omega^{IJ}_{\nu]}
+\omega^{IK}_{[\mu}\omega^{~~~J}_{\nu]K}$ is the field strength of 
the gauge connection $\omega_{\mu}^{IJ}$ corresponding to the local 
SO(3,1) Lorentz symmetry. The fields carry two kinds of indices: 
$\mu\equiv (t,a)\equiv(t,x,y,z)$ referring to the spacetime 
coordinates and $I\equiv(0,i)\equiv (0,1,2,3)$ to the local 
inertial (Lorentz) frame. Completely antisymmetric symbols 
$\epsilon^{\mu\nu\alpha\beta}$ and $\epsilon^{}_{IJKL}$ take 
constant values $0$ and $\pm 1$ with $\epsilon^{txyz}=+1
=\epsilon_{0123}$. The Euler-Lagrange equations of motion 
obtained by varying the action (\ref{HP}) independently with 
respect to $e_\mu^I$ and $\omega_\mu^{IJ}$  are:
\begin{eqnarray}\label{eom1}
\frac{\delta S}
{\delta\omega_{\mu}^{IJ}}~&:&~~~~\epsilon^{\mu\nu\alpha\beta}
\epsilon_{IJKL} e_\mu^K D_{\nu}(\omega)e_{\alpha}^L=0\\
\frac{\delta S}{\delta 
e_{\mu}^{I}}~&:&~~~~\epsilon^{\mu\nu\alpha\beta}\epsilon_{IJKL}
e_\nu^J R_{\alpha\beta}^{~~KL}(\omega)=0\label{eom2}
\end{eqnarray}
This set of equations admits both invertible and noninvertible tetrads 
as solutions. These reduce to Einstein's equations in vacuum only in 
the invertible phase, where the vanishing of torsion emerges as a 
dynamical consequence. For noninvertible tetrads, however, the 
space of solutions consists of geometries that generically exhibit 
torsion even in vacuum \cite{tseytlin,kaul,kaul1}.

Here we attempt to construct a special class of spherically symmetric 
solutions of the equations of motion in pure gravity, which are 
characterized by the  different phases of first order gravity in 
two different regions, one with non-degenerate tetrads and other 
with degenerate tetrads. In particular, we look for spacetime 
solutions with the exterior Schwarzschild metric in one region and 
a degenerate metric in the other. In addition, we demand that these 
must be associated with field-strength whose  components  do not 
diverge anywhere in the manifold and satisfy certain continuity 
properties at the boundary connecting the two regions.
 
Let us note that constructions similar in spirit to the ones discussed 
above have been attempted earlier \cite{bengtsson1,bengtsson,madhavan}. 
For example, Bengtsson \cite{bengtsson,bengtsson1} has presented some   
spacetime solutions of the complex SU(2) formulation 
\cite{sen,ashtekar} of gravity theory with degenerate spatial 
(densitized) triads in the interior. 
 In the explicit examples of real solutions that we 
shall exhibit here, the metrics, while being degenerate in a region, 
are associated with {\it invertible triads}.
 
In the next couple of sections, we present the construction of a class 
of vacuum solutions of first order gravity which exhibit the properties 
outlined above. There are a countably infinity of them, for each 
of which underlies a regular geometry everywhere. The concluding section 
contains  a summary of the main results and a few observations regarding 
the possible importance of these newly found configurations in generic 
contexts.

\section{Region-I: Invertible tetrad}
Let us first introduce a system of coordinates $(t,u,\theta,\phi)$ which 
cover the whole spacetime, with $\left(t\in (-\infty,\infty),~u\in 
(-\infty,\infty),~\theta\in[0,\pi],~\phi\in[0,2\pi]\right)$. In these 
coordinates, we define a spherically symmetric and static metric of the 
form \cite{bengtsson}:
 \begin{eqnarray}\label{sch1}
ds^2=-\left[1-\frac{2M}{f(u)}\right]dt^2 +
\left[1-\frac{2M}{f(u)}\right]^{-1} f'^{2}(u) du^2 + 
f^2(u)\left[d\theta^2+\sin^2 \theta d\phi^2\right]
\end{eqnarray}
where, the monotonic function $f(u)$, which represents the radius of the 
two-sphere (at any fixed $t$ and $u$), satisfies the following properties:
\begin{eqnarray}\label{f(u)}
f(u_0)=2M,~~~~f'(u_0)=0~.
\end{eqnarray}
The metric (\ref{sch1}) describes only a part  of the full 
spacetime (region-I), corresponding to the values $u>u_0$ or $f(u)>2M$. 
The boundary $u=u_0$ represents a three-surface on which the metric 
determinant, $g=-f^4 (u) f^{'2}(u)\sin^2\theta$, vanishes. The constant 
parameter $M$ defines the area of the surface of the two sphere 
$S^2_{(\theta,\phi)}$ at $u=u_0$. For $f(u)>2M$, this metric is a vacuum 
solution of Einstein's equations $R_{\mu\nu}=0$, which essentially 
corresponds to the phase with invertible metrics in first order gravity 
theory. 

The tetrad fields in this region with $f(u)>2M$ can be read off from the 
metric (\ref{sch1}) as:
\begin{eqnarray}\label{tetrad}
e^0_t=\left(1-\frac{2M}{f(u)}\right)^{\frac{1}{2}},~ e^1_u=\frac{f'(u)}
{\left(1-\frac{2M}{f(u)}\right)^{\frac{1}
{2}}},~e_\theta^2=f(u),~e^3_\phi=f(u)\sin\theta
\end{eqnarray} 
The nonvanishing components of the associated (torsionless) spin-connection 
fields $\omega_\alpha^{~IJ}$ are given by:
\begin{eqnarray}\label{omega}
\omega^{01}=\frac{M}{f^2(u)}dt,~\omega^{12}=-\left(1-\frac{2M}
{f(u)}\right)^{\frac{1}{2}} d\theta,~\omega^{23}=-\cos\theta d\phi,~
\omega^{31}=\left(1-\frac{2M}{f(u)}\right)^{\frac{1}{2}}\sin\theta 
d\phi\nonumber\\
\end{eqnarray}
Using these, the field strength tensors $R_{\mu \nu}^{~~IJ}(\omega)$ can be  
evaluated to be:
\begin{eqnarray}\label{R}
R^{01}(\omega)&=&\frac{2Mf'(u)}{f^3(u)}dt\wedge du,~~~R^{02}
(\omega)=-\frac{M}{f^2(u)}\left(1-\frac{2M}{f(u)}\right)^{\frac{1}
{2}}dt\wedge d\theta,\nonumber\\~~R^{03}(\omega)&=&-\frac{M}
{f^2(u)}\left(1-\frac{2M}{f(u)}\right)^{\frac{1}{2}}
\sin\theta ~dt\wedge d\phi,
R^{12}(\omega)=-\frac{Mf'(u)}{f^2(u)}\left(1-\frac{2M}
{f(u)}\right)^{-\frac{1}{2}}du\wedge d\theta,\nonumber\\~R^{23}(\omega)
&=& \frac{2M}{f(u)}\sin\theta ~d\theta\wedge d\phi,~~~R^{31}
(\omega)=-\frac{Mf'(u)}{f^2(u)}\left(1-\frac{2M}{f(u)}
\right)^{-\frac{1}{2}}\sin\theta~ d\phi\wedge du
\end{eqnarray}

From the above set of fields, let us now construct their counterparts in 
the metric formulation,
 namely the affine connection $\Gamma_{\alpha\beta\rho}\equiv 
\Gamma_{\alpha\beta}^{~~\sigma}g_{\rho\sigma}$ and the spacetime field 
strength $R_{\mu\nu\rho\sigma}$, which are invariant under the internal 
$SO(3,1)$ rotations. 
Using the covariant constancy of the tetrad, given by the condition 
${\cal D}_{\mu} e_{\nu}^I \equiv \del_{\mu} e_{\nu}^I 
+\omega_{\mu}^{~IJ}e_{\nu J}-\Gamma_{\mu\nu}^{~~\rho}e_{\rho}^I=0$, 
the affine connection components are given in terms of the basic 
fields ($e_\mu^I,~\omega_\mu^{~IJ})$ as:
\begin{eqnarray}\label{gamma1}
\Gamma_{\mu\nu\rho}=e_{\rho I} \left[\del_\mu 
e_\nu^I+\omega_\mu^{~IJ}e_{\nu J}\right]
\end{eqnarray}
Its nontrivial components are displayed below:
\begin{eqnarray}\label{gamma2}
\Gamma_{ttu}&=&\frac{Mf'(u)}{f^2(u)},~\Gamma_{tut}=-\frac{Mf'(u)}
{f^2}=\Gamma_{utt},~\Gamma_{uuu}=\frac{1}{2}\del_u 
\left[\frac{f(u)f^{'2}(u)}{f-2M}\right],\nonumber\\
\Gamma_{\theta\theta u}&=&-f(u)f'(u),
~\Gamma_{u\theta\theta}=f(u)f'(u)=\Gamma_{\theta u\theta},\nonumber\\
\Gamma_{\phi\phi u}&=&-f(u)f'(u) \sin^2 \theta,~
\Gamma_{u\phi\phi}=f(u)f'(u)\sin^2 \theta=\Gamma_{\phi u\phi},\nonumber\\
\Gamma_{\phi\phi\theta}&=&-
f^2(u)\sin\theta\cos\theta,~\Gamma_{\theta\phi\phi}=f^2 (u)\sin\theta 
\cos\theta=\Gamma_{\phi \theta\phi}~.
\end{eqnarray}
 The tensor $R_{\mu\nu\rho\sigma}$ is defined in terms of the $SO(3,1)$ 
field-strength as:
\begin{eqnarray}\label{curvature}
R_{\mu\nu\rho\sigma}=R_{\mu \nu}^{~IJ}(\omega) e_{\rho I} e_{\sigma J}~,
\end{eqnarray}
whose nonvanishing components read:
\begin{eqnarray}\label{riemann1}
R_{tutu}=-\frac{2Mf^{'2}(u)}{f^3(u)},~R_{t\theta t\theta}=\frac{M[f(u)-2M]}
{f^2(u)},~R_{t\phi t\phi}=\frac{M[f(u)-2M]}{f^2(u)} \sin^2 \theta,
\nonumber\\
R_{u\theta u\theta}=-\frac{Mf^{'2}(u)}{[f(u)-2M]},~R_{\theta\phi 
\theta\phi}=2Mf\sin^2 \theta,~R_{\phi u \phi u}=-\frac{Mf^{'2}(u)}
{[f(u)-2M]} \sin^2 \theta ~.
\end{eqnarray}
In the region $u>u_0$ where tetrad is invertible and torsion is absent, 
the field strength tensor (\ref{curvature}) reduces to the Riemann 
curvature tensor. However, this equality need not hold in general.

\subsection*{Choice of f(u) and boundary conditions:}
Let us note that for $f(u)> 2M$, the metric (\ref{sch1}) is equivalent to the  exterior Schwarzschild solution upto a coordinate transformation. This becomes evident upon the reparametrization
$f(u)=r$, which brings this metric to the Schwarzschild form:
\begin{eqnarray}\label{sc}
ds^2=-\left[1-\frac{2M}{r}\right]dt^2 + \left[1-\frac{2M}{r}\right]^{-1} 
dr^2 + r^2\left[d\theta^2+\sin^2 \theta d\phi^2\right]
\end{eqnarray}
where $r$ is the radial coordinate. However, these two geometries are not 
equivalent at the degenerate surface $u=u_0$, where the coordinate 
transformation defined above becomes ill-defined.

Although $f(u)$ can be any monotonic function which obeys (i) the 
boundary conditions (\ref{f(u)}) and (ii) is such that it does not 
lead to divergences in any of the fields introduced above, it
could nevertheless be more illuminating to work 
with a specific choice for $f(u)$. We choose:
\begin{eqnarray}\label{fchoice}
f(u)&=&2M\left[1+\left(\frac{u}{u_0}-1\right)^{n+1}\right] \label{fchoice1}
\end{eqnarray}
where $n\geq 2$ is an integer. For these values of $n$, this function 
satisfies the above conditions (i) and (ii). At the degenerate boundary 
$u=u_0$, for the explicit choice (\ref{fchoice}), the nonvanishing 
components of the affine connection $\Gamma^{}_{\mu\nu\rho}$ 
in (\ref{gamma2})  and $R_{\mu\nu\alpha\beta}$ tensor in (\ref{riemann1})
exhibit the following behaviour:
  \begin{eqnarray}\label{R1}
&&\Gamma_{\phi\phi\theta}= -\Gamma_{\theta\phi\phi}=-\Gamma_{\phi 
\theta\phi} \doteq -4M^2\sin\theta\cos\theta~;\nonumber\\
 &&R_{\theta\phi \theta\phi}\doteq 4M^2\sin^2 \theta~.
\end{eqnarray}
where the symbol $\doteq$ denotes equality only at $u=u_0$.

The set of fields constructed above defines the vacuum spacetime in 
the region-I,  $u>u_0$ ($f(u)>2M$),  completely. The analysis 
for the other region $u\leq u_0$ is presented next.

\section{Region-II: Noninvertible tetrad}
As emphasized already, our purpose here is to construct degenerate 
spacetime solutions of the first order equations of motion (\ref{eom1}) 
and (\ref{eom2}) in the region  $u\leq u_0$ (region-II). 
To begin with, we consider a degenerate metric with $g_{tt}=0$ 
everywhere in this region:
\begin{eqnarray}\label{metric2}
\hat{ds}^2_{(4)}=0+ \sigma F^2(u) du^2 + H^2(u)\left[d\theta^2+\sin^2 
\theta d\phi^2\right]
\end{eqnarray}
The nondegenerate 3-subspace of this metric exhibits the topology 
$S^2\times R$, which is the same as that of any $t=const.$ slice of the 
metric (\ref{sch1}) in region-I. 
The two possible values of $\sigma=\pm 1$ correspond to an Euclidean or 
a Lorentzian 3-subspace,  respectively. For $\sigma=+1$, $u$ is a 
spacelike coordinate whereas for $\sigma=-1$, it is timelike (in 
region-II). The continuity of the metric requires that the two arbitrary 
functions $F(u)$ and $H(u)$, whose precise forms are to be determined 
using the equations of motion, should have the following behaviour at 
the degenerate boundary:
\begin{eqnarray}
F(u_0)=0,~~~~H(u_0)=2M~.
\end{eqnarray}
The internal (Lorentzian) metric is given by: $\eta_{IJ}=diag[-\sigma,
\sigma,1,1]$. 
 The tetrad fields are:
 \begin{eqnarray}\label{e}
\hat{e}^I_\mu~=~\left(\begin{array}{cccc}
0 & 0 & 0 & 0\\
0 & F(u) & 0 & 0\\
0 & 0 & H(u) & 0\\
0 & 0 & 0 & H(u)\sin\theta \end{array}\right) ~=~ \left(\begin{array}{cc}
0 & 0 \\
0 & \hat{e}_a^i\end{array}\right)
\end{eqnarray}
 The only non-vanishing components of the torsionless spin-connection 
fields 
$\bar{\omega}_a^{~ij}(\hat{e})={\frac{1}{ 2}} \left[ \hat{e}^b_i 
\del^{}_{[a}\hat{e}_{b]}^j-\hat{e}^b_j\del^{}_{[a}\hat{e}_{b]}^i -  
\hat{e}_a^l \hat{e}^b_i\hat{e}^c_j\del^{}_{[b}\hat{e}_{c]}^l \right]$, 
which are determined completely by the triads $\hat{e}_a^i$, are 
given by:
\begin{eqnarray}\label{R3}
\bar{\omega}_\theta^{12}=-\sigma\frac{H'(u)}
{F(u)},~\bar{\omega}_\phi^{23}=-\cos\theta,~\bar{\omega}_\phi^{31}
=\sigma\frac{H'(u)}{F(u)}\sin\theta
\end{eqnarray}
The corresponding $SO(3,1)$ field strength components read: 
\begin{eqnarray}
&&\bar{R}_{u\theta}^{~12}(\bar{\omega})=-\sigma\left[\frac{H'(u)}
{F(u)}\right]',~\bar{R}_{\theta\phi}^{~23}
(\bar{\omega})=\left[1-\sigma\left(\frac{H'(u)}
{F(u)}\right)^2\right]\sin\theta,~\bar{R}_{\phi u}^{~31}(\bar{\omega})=-
\sigma\left[\frac{H'(u)}{F(u)}\right]'\sin\theta\nonumber\\
~
\end{eqnarray}

Given the tetrad fields (\ref{e})  above, we now look for the most 
general set of connection fields $\hat{\omega}_{\mu}^{IJ}\equiv 
(\hat{\omega}_{t}^{0i},~\hat{\omega}_{t}^{ij},~\hat{\omega}_{a}^{ij},
~\hat{\omega}_{a}^{0i})$ which solve  the   
equations of motion (\ref{eom1}). Using the fact that 
the components $\hat{\omega}_a^{ij}$ can be written as a sum of 
the connection $\bar{\omega}_a^{ij}(\hat{e})$ without torsion and 
the contortion $K_a^{ij}$:
\begin{eqnarray}
\hat{\omega}_{a}^{ij}=\bar{\omega}_a^{ij}(\hat{e})+K_a^{ij}~,
\end{eqnarray} 
the most general solution of the equations of motion (\ref{eom1}) is then 
given by \cite{kaul}:
\begin{eqnarray}\label{K}
K_a^{ij}=\epsilon^{ijk} \hat{e}_{a}^l 
N_{kl},~\hat{\omega}_{a}^{0i}=\hat{e}_{al} 
M^{il},~\hat{\omega}_{t}^{0i}=0=\hat{\omega}_{t}^{ij}~,
\end{eqnarray}
where the spacetime-dependent matrices $N_{kl}=N_{lk}$ and $M_{kl}=M_{lk}$ 
are symmetric but   arbitrary otherwise.
The existence of these arbitrary fields is essentially a reflection of 
the fact that in first order gravity theory with noninvertible tetrads, 
the equations of motion leave some of the 
connection components completely undetermined. 
In what follows next, we will restrict our attention to the simpler
case with $M_{kl}=0$. The remaining set of six fields $N_{kl}$ can be
represented as:
\begin{eqnarray}\label{N}
 N_{ij}~=~\left(\begin{array}{ccc}
\alpha & \eta_3 & \eta_2\\
\eta_3 & \beta & \eta_1\\
\eta_2 & \eta_1 & \gamma\end{array}\right) 
\end{eqnarray}
 Using this parametrization of $N^{}_{ij}$  and the triads (\ref{e}), 
 the components of the contortion $K_a^{ij}$ as in (\ref{K}) become:
\begin{eqnarray}
K^{12}&=&\eta_2 F(u)du+H(u)\left[\eta_1 d\theta+\gamma\sin\theta 
d\phi\right]\nonumber\\
K^{23}&=&\alpha F(u)du +H(u)\left[\eta_3 d\theta+\eta_2\sin\theta 
d\phi\right]\nonumber\\
K^{31}&=&\eta_3 F(u)du +H(u)\left[\beta d\theta+\eta_1\sin\theta 
d\phi\right]
\end{eqnarray}
With these, the full connection coefficients $\hat{\omega}^{IJ}$ are 
given by:
\begin{eqnarray}\label{omegafull}
\hat{\omega}^{01}&=& \hat{\omega}^{02}=\hat{\omega}^{03}=0~,\nonumber\\
\hat{\omega}^{12}&=&\eta_2 F(u)du+\left[\eta_1 H(u)-\sigma\frac{H'(u)}
{F(u)}\right]d\theta+\gamma H(u)\sin\theta d\phi~,\nonumber\\
\hat{\omega}^{23}&=&\alpha F(u)du +\eta_3 H(u) d\theta+\left[\eta_2 
H(u)\sin\theta-\cos\theta\right] d\phi~,\nonumber\\
\hat{\omega}^{31}&=&\eta_3 F(u)du +\beta H(u) d\theta+\left[\eta_1 
H(u)+\sigma\frac{H'(u)}{F(u)}\right]\sin\theta d\phi~.
\end{eqnarray}
For these connection fields, the field-strength can be evaluated to be:
\begin{eqnarray}\label{RK}
\hat{R}^{01}(\hat{\omega})&=& \hat{R}^{02}(\hat{\omega})=\hat{R}^{03}
(\hat{\omega})=0~,\nonumber\\
\hat{R}^{12}(\hat{\omega})&=& Fd\eta_2\wedge du+d\left(\eta_1 
H\right)\wedge d\theta
+\sin\theta d(\gamma H)\wedge d\phi 
+\left[(\eta_3^2-\alpha\beta)HF -\sigma\left(\frac{H'}
{F}\right)'\right]du\wedge d\theta\nonumber\\
&+&\left[\gamma H \cos\theta- \eta_3 H\left(\eta_1 H+\sigma\frac{H'}
{F}\right)\sin\theta+\beta H (\eta_2 
H\sin\theta-\cos\theta)\right]d\theta\wedge d \phi\nonumber\\
&+&\left[\alpha F\left(\eta_1 H+\sigma\frac{H'}{F}\right)\sin\theta-\eta_3 
F(\eta_2 H\sin\theta-\cos\theta)\right]d\phi\wedge du~,\nonumber\\
\hat{R}^{23}(\hat{\omega})&=& F d\alpha\wedge du+d(\eta_3 H)\wedge d\theta
+\sin\theta d(\eta_2 H)\wedge d\phi +
\left[\eta_3 H'+\sigma(\eta_2 \beta-\eta_1 \eta_3) FH\right]du\wedge 
d\theta\nonumber\\
&+&\left[\eta_2 H\cos\theta+\sin\theta- \sigma\gamma\beta 
H^2\sin\theta+\sigma\left(\eta_1 H-\frac{H'}{F}\right)\left(\eta_1 
H+\frac{H'}{F}\right)\sin\theta\right]d\theta\wedge d\phi\nonumber\\
&+&\left[-\eta_2 H'+\sigma(\eta_3\gamma-\eta_1 \eta_2) FH\right]\sin\theta 
d\phi\wedge du~,\nonumber\\
\hat{R}^{31}(\hat{\omega})&=& F d\eta_3\wedge du+d(\beta H)\wedge d\theta
+\sin\theta d(\eta_1 H)\wedge d\phi +
\left[-\sigma\alpha H'+(\alpha \eta_1-\eta_2\eta_3) FH \right]du\wedge 
d\theta\nonumber\\
&+&\left[\gamma\eta_3 H^2 \sin\theta+\left(\eta_1 H+\sigma\frac{H'}
{F}\right)\cos\theta-\left(\eta_1 H-
\sigma\frac{H'}{F}\right)\left(\eta_2 H 
\sin\theta-\mathrm{cos}\theta\right)\right]d\theta\wedge d\phi\nonumber\\
&+&\left[\left((\eta_2^2-\alpha\gamma) FH -\sigma\left(\frac{H'}
{F}\right)'\right)\sin\theta- \eta_2 F \cos\theta\right]d\phi\wedge du~.
\end{eqnarray}
This in turn leads to the following identity:
\begin{eqnarray}
\hat{e}_{u}^{1} \hat{R}_{\theta\phi}^{~23}
(\hat{\omega})+\hat{e}_{\theta}^{2} \hat{R}_{\phi u}^{~31}
(\hat{\omega})+\hat{e}_{\phi}^{3} \hat{R}_{u\theta}^{~12}
(\hat{\omega})&=&\left[(\sigma\eta_1^2+\eta_2^2+\eta_3^2-\alpha\beta-\sigma\beta\gamma-
\gamma\alpha)FH^2\right]\sin\theta\nonumber\\
&+&\left[\left(1-\sigma\frac{H'^2}{F^2}\right)F-2\sigma H\left(\frac{H'}
{F}\right)'\right]\sin\theta
\end{eqnarray}

Following \cite {kaul}, it is straight forward to check that the 
configuration described above satisfy the remaining set of equations 
of motion  (\ref{eom2}) provided the contortion fields are 
constrained as: 
\begin{eqnarray}\label{master}
\left(\sigma\eta_1^2+\eta_2^2+\eta_3^2-\alpha\beta-\sigma\beta\gamma-
\gamma\alpha\right)FH^2
+\left(1-\sigma\frac{H'^2}{F^2}\right)F-2\sigma H\left(\frac{H'}
{F}\right)'=0
\end{eqnarray}
Hence, the set of degenerate tetrad (\ref{e}) and the connection fields 
(\ref{omegafull}), subject to the above constraint, solves both the 
equations of motion (\ref{eom1}) and (\ref{eom2}) of first order 
gravity theory in vacuum. These define the geometry of region II.

\section{Joining regions I and II: Full spacetime Solution(s)}
Let us now construct a complete solution for $-\infty<u<\infty$ by 
finding the explicit functional forms of the torsional fields as well 
as of $F(u)$ and $H(u)$, such that the constraint (\ref{master}) is 
obeyed and all the components of the metric, affine connection and 
the field strength are continuous across the $u=u_0$ hypersurface 
which connects regions I and II. Just for the sake of simplicity, 
in the rest of our analysis we choose a simpler setting where only 
one of the six torsional fields is nonvanishing and depends only on $u$:
\begin{eqnarray}\label{choice}
\eta_1=\eta (u),~\eta_2=0,~\eta_3=0,~\alpha=0,~\beta=0,~
\gamma=0.
\end{eqnarray}
For this choice, the nonvanishing components of the affine connection   
$\hat{\Gamma}_{\mu\nu\rho}=\hat{e}_{\rho I} \left[\del_\mu 
\hat{e}_\nu^I+\hat{\omega}_\mu^{~IJ}\hat{e}_{\nu J}\right]$ 
(which contain torsion now) are evaluated to be:
\begin{eqnarray}\label{gamma3}
\hat{\Gamma}_{uuu}&=&\sigma F(u) F'(u),~
\hat{\Gamma}_{\theta\theta u}=\sigma 
H(u)F(u)\left(\eta(u)H(u)-\sigma\frac{H'(u)}
{F(u)}\right)=-\hat{\Gamma}_{\theta u\theta} 
,~\hat{\Gamma}_{u\theta\theta}=H(u)H'(u),\nonumber\\
\hat{\Gamma}_{\phi\phi u}&=&-\sigma 
H(u)F(u)\left(\eta(u)H(u)+\sigma\frac{H'(u)}{F(u)}\right) \sin^2 
\theta=-\hat{\Gamma}_{\phi u\phi},
~\hat{\Gamma}_{u\phi\phi}=H(u)H'(u)\sin^2 \theta,\nonumber\\
\hat{\Gamma}_{\phi\phi\theta}&=&-
H^2(u)\sin\theta\cos\theta=-\hat{\Gamma}_{\phi\theta\phi}
=-\hat{\Gamma}_{\theta\phi\phi}~.
\end{eqnarray}
The spacetime field strength 
$\hat{R}_{\mu\nu\alpha\beta}=\hat{R}_{\alpha\beta}^{~IJ} 
(\hat{\omega})\hat{e}_{\mu I} \hat{e}_{\nu J}$ in this case becomes:
\begin{eqnarray}\label{R-simple}
\hat{R}_{u\theta u \theta}&=&\sigma\left(\eta (u) H(u)-\sigma\frac{H'(u)}
{F(u)}\right)'F(u) H(u) ~,\nonumber\\
\hat{R}_{\theta\phi\theta\phi}&=&\left[1+\sigma\left(\eta(u) 
H(u)-\frac{H'(u)}{F(u)}\right)\left(\eta(u) H(u)+\frac{H'(u)}
{F(u)}\right)\right]H^2(u) \sin^2 \theta~, \nonumber\\
\hat{R}_{\phi u\phi u}&=&-\sigma\left(\eta(u) H(u)+\sigma\frac{H'(u)}
{F(u)}\right)'F(u) H(u)\sin^2 \theta 
\end{eqnarray}
while all the other components turn out to be zero. In particular, the 
components $\hat{R}_{tutu},~\hat{R}_{t\theta t\theta}$ and 
$\hat{R}_{t\phi t\phi}$ vanish   everywhere in the region-II ($u\le u_0$).
This is to be contrasted with the behaviour in region-I where the 
field strength components  $R_{tutu}, ~R_{t\theta t\theta}$ and 
$R_{t\phi t\phi}$ as presented in (\ref{riemann1}) are zero only 
at the boundary. 
 
For the choice (\ref{choice}),  the constraint (\ref{master})  among 
the fields $\eta(u),H(u),F(u)$ reduces to:
\begin{eqnarray}\label{master1}
\eta^2 FH^2
+\left(\sigma-\frac{H'^2}{F^2}\right)F-2 H\left(\frac{H'}{F}\right)'=0
\end{eqnarray}
This provides only one condition among the three unknown fields. Since 
there are no more equations of motion that could be used to solve for 
these, we have the freedom of choosing two further constraints, such that 
the continuity properties at $u=u_0$ are satisfied. To this end, let us 
consider the following ansatze:
\begin{eqnarray}\label{eqn1}
\eta(u)&=&\sigma\frac{H'(u)}{F(u)H(u)},\nonumber\\
\frac{H'(u)}{F(u)}&=&\beta e^{\alpha(u-u_0)} (u_0-u)^m
\end{eqnarray} 
where $\alpha>0$ and $\beta$ are real-valued constants and $m$ can 
be an integer or a half-integer. 
With this, the components in eq.(\ref{R-simple}) simplify to:
\begin{eqnarray}\label{R-simp}
\hat{R}_{u\theta u \theta}=0,~
\hat{R}_{\theta\phi\theta\phi}= H^2(u) \sin^2 \theta,~
\hat{R}_{\phi u\phi u}=-2\left(\frac{H'(u)}{F(u)}\right)'
F(u) H(u)\sin^2 \theta ~.
\end{eqnarray}

The three equations in (\ref{master1}) and (\ref{eqn1}) now can be 
solved for the three fields $H(u),F(u)$ and $\eta(u)$, leading to:
\begin{eqnarray}\label{nonsingular1}
F(u)&=&4\sigma\beta M \left[\alpha(u_0-u)^m-m(u_0-u)^{m-1}\right] 
~e^{\left[\alpha(u-u_0)+\sigma\beta^2 (u_0-u)^{2m} e^{2\alpha
(u-u_0)}\right]},\nonumber\\ 
H(u)&=&2M ~e^{\left[\sigma\beta^2 (u_0-u)^{2m} 
e^{2\alpha(u-u_0)}\right]},\nonumber\\ 
\eta(u)&=&\sigma \frac{\beta}{2 M} (u_0-u)^m e^{\left[\alpha
(u-u_0)-\sigma\beta^2 (u_0-u)^{2m} e^{2\alpha(u-u_0)}\right]}
\end{eqnarray}
The constant $\alpha$ can be fixed by using the freedom in choosing 
the origin of $u$ coordinate. If we choose $u=0$ to be the point where 
the radius of the two sphere ($H(u)$) is extremum, then we have 
$\alpha=\frac{m}{u_0}$ for a fixed $m$.
The requirement of continuity of the metric components $g_{\mu\nu}$ 
at $u=u_0$ fixes the other constant $\beta$ along with the number $m$ 
as:
\begin{eqnarray*}
\beta=u_0^{-m},~~~~~m=\frac{n+1}{2}
\end{eqnarray*}
where $n\geq 2$ is the same integer appearing in the definition of $f(u)$ 
in eq.(\ref{fchoice}). This leads to two sets of solutions. The first 
one, corresponding to $\sigma=+1$, are represented by the following 
fields:
\begin{eqnarray}\label{nonsingular2}
F(u)&=&-\frac{4 M}{u_0^2} p u \left(1-\frac{u}{u_0}\right)^{p-1} 
e^{\left[p\left(\frac{u}{u_0}-1\right)+ \left(\frac{u}{u_0}-1\right)^{2p} 
e^{2p\left(\frac{u}{u_0}-1\right)}\right]},\nonumber\\
H(u)&=&2M ~e^{\left[ \left(\frac{u}{u_0}-1\right)^{2p} 
e^{2p\left(\frac{u}{u_0}-1\right)}\right]},\nonumber\\ 
\eta(u)&=& \frac{1}{2 M} \left(1-\frac{u}{u_0}\right)^p 
e^{\left[p\left(\frac{u}{u_0}-1\right)- 
\left(\frac{u}{u_0}-1\right)^{2p} e^{2p\left(\frac{u}{u_0}
-1\right)}\right]}~,
\end{eqnarray}
where $p=\frac{n+1}{2}\geq 2$ is an integer. The other set with 
$\sigma=-1$ is given by:
\begin{eqnarray}\label{nonsingular1}
F(u)&=&\frac{4  M}{u_0^2} l u \left(1-\frac{u}{u_0}\right)^{l-1} 
e^{\left[l\left(\frac{u}{u_0}-1\right)+ 
\left(\frac{u}{u_0}-1\right)^{2l} e^{2l\left(\frac{u}
{u_0}-1\right)}\right]},\nonumber\\
H(u)&=&2M ~e^{\left[ \left(\frac{u}{u_0}-1\right)^{2l} e^{2l
\left(\frac{u}{u_0}-1\right)}\right]},\nonumber\\ 
\eta(u)&=& -\frac{1}{2 M} \left(1-\frac{u}{u_0}\right)^l 
e^{\left[l\left(\frac{u}{u_0}-1\right)- 
\left(\frac{u}{u_0}-1\right)^{2l} e^{2l\left(\frac{u}
{u_0}-1\right)}\right]}~,
\end{eqnarray}
where $l=\frac{n+1}{2}\geq \frac{3}{2}$ is a half-integer.
With this, we have a countable infinity of vacuum solutions of the 
first order equations of motion, parametrized by the integer $n\geq 2$, 
whose odd or even values correspond to $\sigma=+1$ and $\sigma=-1$, 
respectively. The metric components $g_{\mu\nu}(u)$ are $C^{n}$ functions. 
Note that  the parameter $M$ has the interpretation of being the inverse 
of the contortion $\eta$ at the extremal point $u=0$ upto a 
numerical constant that depends on $n$.
 
For the solutions displayed above, the   nonvanishing components of 
the affine connection and the field strength at the boundary $u=u_0$ 
are given by:
\begin{eqnarray}\label{gamma4}
 \hat{\Gamma}_{\phi\phi\theta}&=&-\hat{\Gamma}_{\phi\theta\phi}
=-\hat{\Gamma}_{\theta\phi\phi}\doteq -4M^2 \sin\theta 
\cos\theta~;\nonumber\\
\hat{R}_{\theta\phi \theta\phi}&\doteq& 4M^2\sin^2 \theta~.
\end{eqnarray}
Comparing of these with the corresponding boundary values (\ref{R1}) 
for the nondegenerate region, we note that the set of $SO(3,1)$ invariant 
fields are all continuous at $u=u_0$:
\begin{eqnarray}\label{cont1}
g_{\rho\sigma}\doteq 
\hat{g}_{\rho\sigma},~\Gamma_{\rho\sigma\alpha}\doteq
\hat{\Gamma}_{\rho\sigma\alpha},~
R_{\alpha\beta\rho\sigma}\doteq \hat{R}_{\alpha\beta\rho\sigma}~.
\end{eqnarray}

Of the original $SO(3,1)$ valued fields, while all the tetrad and field
strength components are continuous across the separating degenerate 
boundary, some connection components, which are pure gauge on 
the boundary, are not continuous. 
But all the $SO(3,1)$ invariant fields as reflected in (\ref{cont1}) 
are continuous across this  boundary.

Let us look at the nature of  geometries in the  region-II as represented
by  the fields given in eqs.(\ref{nonsingular2}) and (\ref{nonsingular1}) 
in detail.  For both these sets, the fields have the following boundary 
behaviour:
\begin{eqnarray}
F(u)\rightarrow 0 , ~H(u)\rightarrow 2M,~\eta(u)\rightarrow 0 
~\mathrm{~~as} ~u\rightarrow -\infty~\mathrm{or ~}u\rightarrow u_0.
\end{eqnarray} 
Note that the radius  $H(u)$ of the two-sphere is finite and nonvanishing 
everywhere. For $\sigma=+1$, it has a  maximum value at $u=0$, given by 
$H_{max}=2M \exp(e^{-2p})>2M$ (for a fixed integer $p\geq 2$). On the 
other hand, for $\sigma=-1$, it exhibits a  minimum value at $u=0$, 
with $H_{min}=2M \exp(-e^{-2l})<2M$ (for a fixed half-integer 
$l\geq \frac{3}{2}$). This has been displayed in the Fig.1 where 
the profile of the radius of two-sphere $R(u) = f(u)$ for 
the region-I ($u > u_0$) and $R(u) = H(u)$ for the region-II ($u\le u_0$) 
has been presented. The interpretation of the two free parameters $M$ 
and $u_0$ in each solution is now apparent: they characterize the  area and location of the hypersurface at $u=u_0$, respectively.
\begin{figure}\begin{center}
\includegraphics[height=14cm]{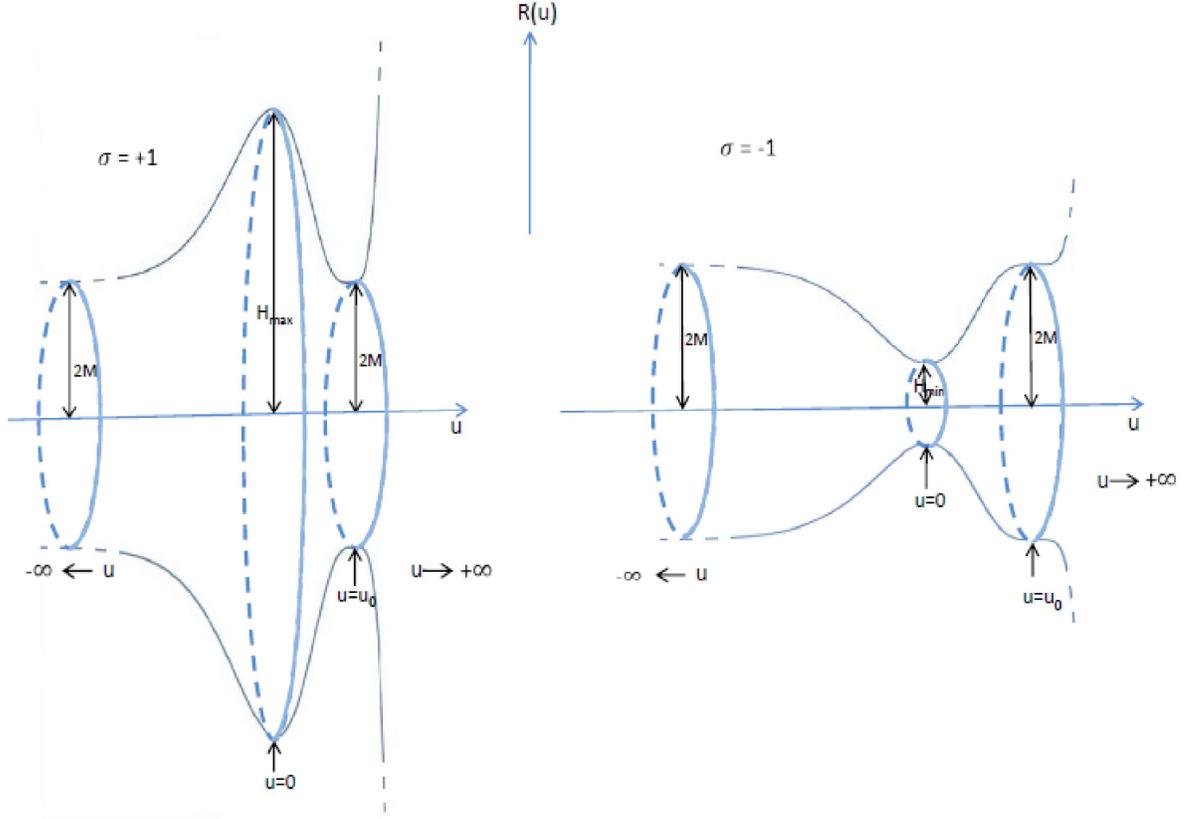}
\caption{Pictorial representation of spacetime solutions for 
$\sigma=\pm 1$ at any fixed $t$}
\end{center}
\end{figure}
 The behaviour of the contortion field $\eta(u)$, which is localized entirely within region-II around the origin $u=0$, has been been provided in Fig.2. The profiles for $\sigma=\pm 1$ are qualitatively the same (for any fixed integer $p\geq 2$ or any fixed half integer $l\geq \frac{3}{2}$), the extrema being at $u=0$.
\begin{figure}\begin{center}
\includegraphics[height=11cm]{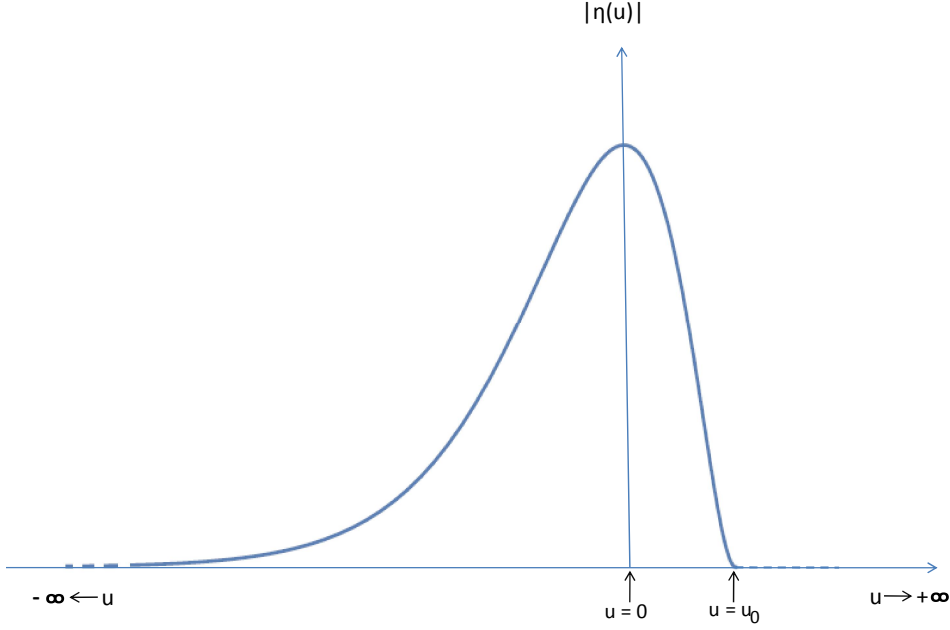}
\caption{Contortion field $\eta(u)$ for $\sigma=\pm 1$}
\end{center}
\end{figure} 
 
It should be emphasized that the spacetime field strength components 
$\hat{R}_{\mu\nu\rho\sigma}$ are finite everywhere in the range
$-\infty<u<\infty$. Although it is not possible to construct  
scalars  from the fields $\hat{R}_{\mu\nu\rho\sigma}$ 
(unless they are topological) associated with a noninvertible 
4-metric, one can nevertheless look at the scalars associated 
with the nondegenerate 3-subspace described by 
$(e_a^i,~\hat{\omega}_b^{kl})$. These  are  well-behaved 
in the entire domain:
\begin{eqnarray*}
&& \hat{R}(\hat{\omega})=\hat{e}^a_i \hat{e}^b_j \hat{R}_{ab}^{~ij}
(\hat{\omega})=0;\\
&& \hat{R}_{ab}^{~ij}(\hat{\omega})\hat{R}^{ab}_{~ij}
(\hat{\omega})=\hat{R}_{u\theta}^{~12}(\hat{\omega})
\hat{R}^{u\theta}_{~12}(\hat{\omega})+\hat{R}_{\theta\phi}^{~23}
(\hat{\omega})\hat{R}^{\theta\phi}_{~23}(\hat{\omega})
+\hat{R}_{\phi u}^{~31}(\hat{\omega})\hat{R}^{\phi u}_{~31}
(\hat{\omega})= \frac{4}{H^4(u)}
\end{eqnarray*}
 
The configurations described above are to be contrasted with 
the Schwarzschild spacetime, which is the unique spherically 
symmetric solution of Einsteinian gravity and is associated 
with divergent field strength components $R_{trtr},~
R_{t\theta t\theta}$ and $R_{t\phi t\phi}$ at the origin.

\section{conclusions}
First order formulation of classical gravity theory in four dimensions 
admits degenerate spacetimes as vacuum solutions. Based on this 
observation, we have  constructed a class of spherically symmetric 
geometries with two regions which are associated with invertible and 
noninvertible tetrads. As the field configurations in 
both these regions satisfy the first order equations of motion in pure
gravity and are continuous across the degenerate boundary connecting 
them, the full spacetime as a whole represents a vacuum solution of 
gravity theory. In the region with non-degenerate metric, away from 
the separating  boundary, the spacetime  geometry is equivalent to  
that of the Schwarzschild exterior.

The most remarkable property of these solutions are reflected through 
the field-strength components, which are well-behaved everywhere. In 
this sense these spacetimes are regular, since any curvature singularity 
is typically a reflection of the divergence in the individual field 
strength components. It should be emphasized that the existence 
of these solutions of the first order equations of motion is 
not in any way in conflict with Birkhoff's theorem, which concerns 
solely the invertible phase ($\det g_{\mu\nu}\neq 0$) of pure gravity.

Within the degenerate region, as described by the associated noninvertible metric, the spacetime essentially becomes two-dimensional at the points $u=u_0$, $u=0$ and also at $u\rightarrow -\infty$ which is one of the asymptotic boundaries. It is not clear at this stage whether such a phenomenon really does encode a change of spacetime topology in classical gravity.

The general framework presented here can also be used to construct vacuum
solutions in first order gravity theory with multiple regions containing
degenerate and non-degenerate geometries. In particular, one such
solution with two regions of flat spacetime separated by a finite
sized bridge which has a non-invertible metric has been presented
in ref \cite{sandipan}.

Finally, let us note that the configurations discussed here 
correspond to finite (vanishing) action.   
Fluctuations around these saddle points might encode nontrivial 
contributions to the path integral of quantum gravity. These 
vacuum geometries may be expected to be relevant in other 
formulations of quantum gravity as well. 

\acknowledgments
Discussions with Amit Ghosh, Ghanashyam Date, Suvrat Raju, Nemani Suryanarayana, Somnath Bharadwaj, Sayan Kar, Soumitra Sengupta and Madhavan Varadarajan, as well as the help of Debraj Choudhury and Sajal Dhara in generating the diagrams are gratefully acknowledged by S.S. He is supported by the grant no. ECR/2016/000027 under the SERB, DST, Govt. of India. R.K. acknowledges the support of DST 
through a J.C. Bose National Fellowship.

\end{document}